\documentclass{ws-ijmpcs}

\begin{document}

\markboth{A. Raponi et al.}
{The Solar Limb Darkening Function With Eclipse Observations}

%
\catchline{}{}{}{}{}
%

\title{ECLIPSES, SOLAR LIMB DARKENING FUNCTION AND DIAMETER MEASUREMENTS: TOWARD A UNIFIED APPROACH}

\author{ANDREA RAPONI}

\address{Sapienza University of Rome, \\P.le Aldo Moro 5 00185, Roma (Italy), \\
andr.raponi@gmail.com}

\author{COSTANTINO SIGISMONDI}

\address{Sapienza University of Rome,\\ 
ICRA, International Center for Relativistic Astrophysics, \\P.le Aldo Moro 5 00185, Roma (Italy)}

\author{KONRAD GUHL, RICHARD NUGENT, ANDREAS TEGTMEIER}

\address{IOTA, International Occultation Timing Association}

\maketitle

\begin{history}
\received{3 Jan 2012}
\revised{Day Month Year}
\end{history}

\begin{abstract}
In order to perform high resolution astrometry of the solar diameter from the ground, through the observations of eclipses, the study of the limb darkening profile is described. Knowing the profile of the solar limb is useful both to monitor the solar radius over time, and to define the oblateness, which is interesting for the classical tests of general relativity. The Limb Darkening Function (LDF) is studied in order to find the inflexion point, to which the measurements of the solar diameter are referred.
The proposed method is applied to the videos of the annular eclipse in January, 15, 2010.
This new method might solve the ambiguity of some eclipse obervations made with different instruments, where the measured solar diameter was clearly dependent on the aperture of the telescope and on the density of the filter used.

\end{abstract}

\ccode{PACS numbers: 95.10.Jk, 95.10.Gi, 96.60.-j, 96.60.Bn}

\section{The method of eclipses}

With the eclipse observation we are able to bypass the atmospheric and instrumental effects that distort the shape of the limb.\cite{Djafer} They are overcome by the fact that the scattering of the Sun's light is greatly reduced by the occultation of the Moon, therefore there are much less photons from the photosphere to be poured, by the PSF effect, in the outer region. 
The observation of the LDF through the observation of eclipses has already had several contributions.\cite{Righini,Rubin,Weart} These early works took into account the nature of the rugged lunar profile, but like a nuisance to be overcome.
A decisive breakthrough was made thanks to David Dunham (IOTA) that proposed to observe the beads of light that appear or disappear from the bottom of a lunar valley
when the solar limb is almost tangent to the lunar limb (Baily's Beads).

\subsection{Comparison between observation and ephemeris} 

It is not the positions of the Baily's Beads to be directly measured, but the timing of appearing or disappearing. In fact the times when the photosphere disappears or emerges behind the valleys of the lunar limb, are determined solely by the positions of the Sun and the Moon, their angular size and lunar profile at the instant, bypassing in this way the atmospheric seeing.

The International Occultation Timing Association (IOTA) is currently engaged to observe the eclipses with the aim of measuring the solar diameter. This is facilitated by the development of the software Occult 4 by David Herald\footnote[1]{$www.lunar-occultations.com/iota/occult4.htm$}.

The technique consists to look at the time of appearance of the beads and to compare it with the calculated positions by the ephemeris using the software Occult 4. The simulated Sun by Occult 4 has the standard radius: 959.63 arcsec at 1AU.\cite{Auwers} The difference between the simulations and the observations is a measure of the radius correction with respect to the standard radius ($\Delta$R).

However this approach conceives the bead as an on-off signal. In this way one assumes the Limb Darkening Function as a Heaviside profile, but it is actually not. Different compositions of optical instruments (telescope + filter + detector) could have different sensitivities and different signal/noise ratios, recording the first signal of the bead in correspondence of different points along the luminosity profile. This leads to different values of the $\Delta$R (see Fig. 1).
An improvement of this approach has to take into account the whole shape of the Limb Darkening Function and thus the actual position of the inflection point.

\begin{figure}
\centerline{\includegraphics[width=0.5\textwidth,clip=]{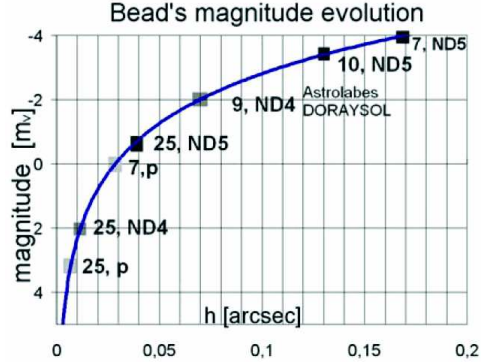}}
\caption{Bead´s magnitude evolution: the height of the solar limb above the valley is on the abscissa. The various square dots represent different type of telescopes. [25, p] means 25 cm opening with projection of the image. [25, ND4] the 25 cm telescope with a filter of Neutral Density of transmittance 1/10000; ND5 stands for 1/100000, and so on.}
\label{Fig. 1}
\end{figure}

\subsection{Numerical calculation}

The shape of the light curve of the bead is determined by the shape of the LDF (not affected by seeing) and the shape of the lunar valley that generates the bead. 
Calling w(x) the width of the lunar valley (i.e. the length of the solar edge visible from the valley in function of the height x from the bottom of the valley), and B(x) the surface brightness profile (i.e. the LDF), one could see the light curve L(y) as a convolution of B(x) and w(x), being $\mid y\mid$ the distance between the botton of the lunar valley and the standard solar edge, setting to 0 the position of the standard edge.

\noindent $L(y)=\intop B(x)\, w(y-x)\, dx$

Thus the profile of the LDF and the lunar valley profile are discretized in order to obtain the solar layers B(n) of equal thickness and concentric to the center of the Sun. In the short space of a lunar valley these layers are roughly parallel and straight. 

\noindent The discrete convolution is:

\noindent $L(m)=\sum B(n)\, w(m-n) h$

\noindent where n, m are the index of the discrete layers corresponding to x, y coordinate and h the layers thickness.

During a bead event every lunar layer is filled by one solar layer every given interval of time. Step by step during an emerging bead event, a deeper layer of the solar atmosphere enters into the profile drawn by the lunar valleys (see Fig. 2), and each layer casts light through the same geometrical area of the previous one. 

\begin{figure}
\centerline{\includegraphics[width=0.7\textwidth,clip=]{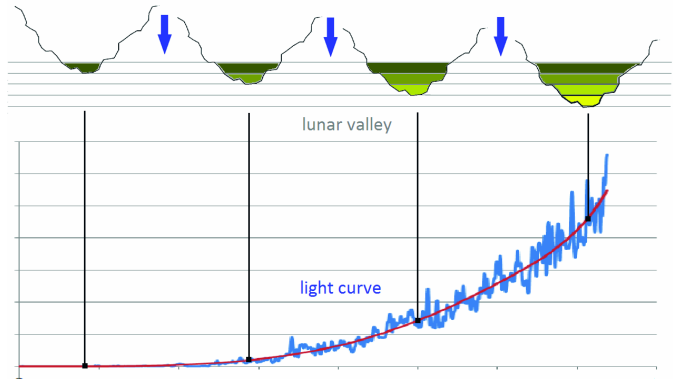}}
\caption{Every step in the geometry of the solar-lunar layers (up) corresponds to a given instant in the light curve (down). The value of the light curve is the contribute of all the layers.}
\label{Fig. 2}
\end{figure}

\begin{figure}
\centerline{\includegraphics[width=0.7\textwidth,clip=]{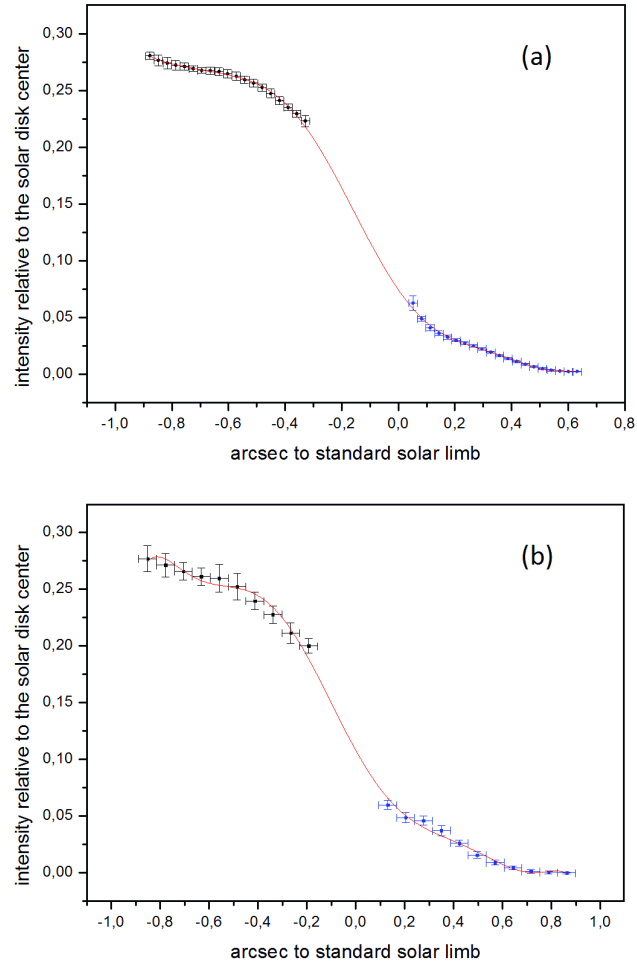}} 
\caption{The luminosity profiles obtained are plotted and put together. The inner and brighter parts are obtained from Tegtmeier's video; the outer and weaker parts are obtained from Nugent's video. Panels (a) and (b) show respectively the luminosity profile concerning the bead at AA = 177$^\circ$ and AA = 171$^\circ$. The luminosity profiles are normalized to the center of the solar disk according to Rogerson.\protect\cite{Rogerson} The zero of the abscissa is the position of the standard solar limb with a radius of 959.63 arcsec at 1 AU. The error bars on y axis are the 90\% confidence level. The error bars on x axis are the thickness (h) of the lunar layers. The solid line is an interpolation between the profiles and gives a possible scenario on the position of the inflection point.}
\label{Fig. 3}
\end{figure}

\noindent The situation described above relates to an emerging bead. The same process can run for a disappearing bead, simply plotting the light curve back in time.

\section{An application of the method}

We studied the videos of the annular eclipse in January 15, 2010 realized by Richard Nugent in Uganda and Andreas Tegtmeier in India with CCD camera Watec and Matsukov telescope.

Two beads located at Axis Angle\footnote[1]{the angle around the limb of the Moon, measured Eastward from the Moon's North pole} (AA) $=171^\circ, 177^\circ$ are analyzed for both the videos. 

The lunar valley analysis is performed with the Occult 4 software that exploits the new lunar profile obtained by the laser altimeter (LALT) onboard the Japanese lunar explorer Kaguya \footnote[3]{$http://wms.selene.jaxa.jp/selene\_viewer/index\_e.html$}. The choice of the thickness of the layers has to be optimal: large enough to reduce $B_n$ uncertainties, but small enough to have a good resolution of the LDF. 

The resulting points show the profiles obtained in two different positions. The inflection point is clearly between the two profiles (see Fig. 3) leading to a value of $-0.19 <\Delta R< +0.05$. 

The same eclipse analyzed by Andreas Tegtmeier and Konrad Guhl led to a value of $\Delta R = -0.29 \pm 0.10$ arcsec.\cite{Guhl} This result doesn't seem compatible with the possible range for the
position of the inflection point we found. This shows that the solar radius defined by the classical method is different than that defined by the inflection point of
the LDF.

\section{Conclusions}

This study takes into account the potentiality of the observation of eclipses in defining the luminosity profile of the edge of the Sun. 

The method proposed considers the bead as a light curve forged by the LDF and the profile of the lunar valley. A first application on two beads of the annular eclipse on 15 January 2010, is described in this study. We obtain a detailed profile, demonstrating the functionality of the method. Although it was impossible to observe the inflection point, its position is defined within a narrow range. The solar radius is thus defined within this range (from -0.19 to +0.05 with respect to the standard radius), resulting compatible with the standard value.



\begin{thebibliography}{0}    

\bibitem{Djafer} D. Djafer, G. Thuillier, S. Sofia, {\it Astrophys. J.} {\bf 676}, 651-657 (2008).

\bibitem{Righini} G. Righini, M. Ballario, G. Godoli, {\it Memorie della Societ\`a Astronomica Italiana} {\bf 24}, 3 (1953).

\bibitem{Rubin} V. C. Rubin, {\it Astrophys. J.} {\bf 129}, 812 (1959).

\bibitem{Weart} S. R. Weart, J. E. Faller, {\it Astrophys. J.} {\bf 157}, 887 (1969).

\bibitem{Auwers} A. Auwers, {\it Astronomische Nachrichten} {\bf 128}, 361 (1891).

\bibitem{Guhl} A. Tegtmeier, K. Guhl, {\it Journal for Occultation Astronomy}, vol. {\bf 1} n. 4 October-December, 19  (2011).

\bibitem{Rogerson} J. B. Rogerson, {\it Astrophys. J.}, {\bf 130}, 985 (1959).


\end{thebibliography}
\end{document}